\documentclass[aps,prd,nopacs,nofootinbib,notitlepage,superscriptaddress,twocolumn]{revtex4-1}

\pdfoutput=1

\usepackage{amsfonts}
\usepackage{amsmath}
\usepackage{xcolor}
\usepackage[colorlinks=true,hyperfootnotes=true,citecolor=cyan]{hyperref}

\newcommand{\diff}{\textrm{d}}

\newcommand{\Q}{\mathbb{Q}}

\newcommand{\mS}{\mathcal{S}}

\newcommand{\mpl}{M_{\text{Pl}}}

\newcommand{\be}{\begin{equation}}
\newcommand{\ee}{\end{equation}}

\newcommand{\lsim}   {\mathrel{\mathop{\kern 0pt \rlap
  {\raise.2ex\hbox{$<$}}}
  \lower.9ex\hbox{\kern-.190em $\sim$}}}
\newcommand{\gsim}   {\mathrel{\mathop{\kern 0pt \rlap
  {\raise.2ex\hbox{$>$}}}
  \lower.9ex\hbox{\kern-.190em $\sim$}}}

\newcommand{\cg}[1]{{\mathring{ #1}}}

\begin{document}

\title{A class of ghost-free theories in symmetric teleparallel geometry}

\author{Antonio G. Bello-Morales}
\email{antgon12@ucm.es}
\address{Departamento de F\'isica Te\'orica and Instituto de F\'isica de Part\'iculas y del Cosmos (IPARCOS-UCM),
Universidad Complutense de Madrid, 28040 Madrid, Spain}

\author{Jose Beltr{\'a}n Jim{\'e}nez}
\email{jose.beltran@usal.es}
\affiliation{Departamento de Física Fundamental and IUFFyM, Universidad de Salamanca, E-37008 Salamanca, Spain}

\author{Alejandro Jim{\'e}nez Cano}
\email{alejandro.jimenez.cano@upm.es}
\address{Laboratory of Theoretical Physics, Institute of Physics, University of Tartu, W. Ostwaldi 1, 50411 Tartu, Estonia}
\address{Escuela T\'ecnica Superior de Ingenier\'ia de Montes, Forestal y del Medio Natural, Universidad Politécnica de Madrid, 28040 Madrid, Spain}

\author{Tomi S. Koivisto}
\email{tomi.koivisto@ut.ee}
\address{Laboratory of Theoretical Physics, Institute of Physics, University of Tartu, W. Ostwaldi 1, 50411 Tartu, Estonia}
\address{National Institute of Chemical Physics and Biophysics, R\"avala pst. 10, 10143 Tallinn, Estonia}

\author{Antonio L. Maroto}
\email{maroto@ucm.es}
\address{Departamento de F\'isica Te\'orica and Instituto de F\'isica de Part\'iculas y del Cosmos (IPARCOS-UCM),
Universidad Complutense de Madrid, 28040 Madrid, Spain}

\begin{abstract}
Theories formulated in the arena of teleparallel geometries are generically plagued by
ghost-like instabilities or other pathologies that are ultimately caused by the breaking of some symmetries. In this work, we construct a class of ghost-free theories based on a symmetry under Transverse Diffeomorphisms that is naturally realised in symmetric teleparallelism. We explicitly show their equivalence to a family of theories with an extra scalar field plus a global degree of freedom and how Horndeski theories and healthy couplings to matter fields can be readily accommodated.
\end{abstract}

\maketitle

\section{Introduction}

The equivalence principle, that can be understood as a consistency requirement for the low-energy interactions of a massless spin-2 particle, leads to a natural interpretation of gravity as the geometry of spacetime. Once the geometrical character of gravity is promoted to a fundamental property, the appropriate geometrical framework needs to be specified. The common approach incepted by Einstein's General Relativity (GR) uses a pseudo-Riemannian geometry where gravity is ascribed to the spacetime metric and the affine structure is completely fixed to be governed by the unique symmetric and metric-compatible connection. This possibility is however not unique and the very same Einstein equations admit equivalent formulations in teleparallel geometries, i.e., geometries with a flat connection. If we constrain the connection to be also metric-compatible (constituting the metric teleparallel geometries), it is possible to construct the (Metric) Teleparallel Equivalent of GR (MTEGR) \cite{Maluf:2013gaa} by requiring a local Lorentz symmetry, while if we impose the connection to be symmetric (symmetric teleparallel geometries), we can formulate a Symmetric Teleparallel Equivalent of GR (STEGR) \cite{Nester:1998mp,BeltranJimenez:2017tkd} by requiring a second copy of Diffeomorphisms (Diffs) invariance. These three equivalent formulations conform the so-called Geometrical Trinity of Gravity \cite{BeltranJimenez:2019esp}. This class of equivalent formulations of gravity can be further extended to the General Teleparallel Equivalent of GR (GTEGR) \cite{BeltranJimenez:2019odq}, where the connection is only required to be flat, but we require an additional local GL(4,$\mathbb{R}$) symmetry.

Having at our disposal the discussed equivalent geometrical formulations of GR, alternative routes to explore extensions are opened. These extensions are however usually problematic already at the theoretical level because they inherently come with the breaking of local symmetries that jeopardise their consistency. The presence of pathologies can be subtle in some cases. For instance, strong coupling problems seem to be ubiquitous in the so-called $f(T)$ theories for many physically relevant backgrounds \cite{Li:2010cg,Ong:2013qja,BeltranJimenez:2020fvy,Bahamonde:2022ohm}, but this does not preclude performing e.g. linear perturbations without encountering any apparent inconsistencies. These strong coupling problems are usually associated with the appearance of accidental local symmetries for the linear spectrum around those backgrounds. The hazards are more obvious in other frameworks such as the symmetric teleparallel geometry where extensions of the STEGR commonly entail introducing derivative self-interactions for the graviton that break the full Diffs invariance and, thus, are prone to propagating Ostrogradski ghosts. This has been put forward in \cite{Gomes:2023tur} where the unavoidable presence of ghosts for the $f(\Q)$ theories has been explicitly unveiled.

In this note we will be concerned with the symmetric teleparallel theories and our aim will be to show that the original symmetries of the STEGR can be minimally broken so as to obtain a class of theories that are manifestly ghost-free. We will see that the constructed theories explicitly break Diffs to its Transverse Diffs (TDiffs) subgroup and that this suffices to guarantee the absence of ghosts at the full non-linear level. 
Incidentally, we will show that these theories are equivalent to a class of scalar field theories with the particularity that they contain a constant in the potential that arises as an integration constant, i.e., it is a global degree of freedom similar to the cosmological constant of unimodular gravity (see e.g. \cite{Carballo-Rubio:2022ofy} and references therein).

The paper is organized as follows: We will commence by reviewing the symmetric teleparallel geometries in Sec.\ref{Sec:SymTel}. We will then move to the main focus of this work in Sec. \ref{Sec:TDiffNGR} where we will consider a TDiff invariant extension of the STEGR. We will show how the Stuckelberg fields can be integrated out and an equivalent scalar field formulation can be constructed. We will briefly discuss the symmetries and the cosmologies of these theories. We will then proceed to the construction of other ghost-free extensions beyond the general quadratic theories in Sec. \ref{Sec:GFExt}. The couplings to matter fields will be discussed in Sec.\ref{Sec:MC} and we will conclude with the discussions in \ref{Sec:Discussion}.

\section{Symmetric Teleparallelism}
\label{Sec:SymTel}

A teleparallel geometry is defined by the requirement of being flat, i.e., the connection must have vanishing curvature $R^\alpha{}_{\beta\mu\nu}=0$. This condition can be integrated for the connection in terms of an arbitrary {\it reference (inertial) frame} $\Lambda^\alpha{}_\mu$ so that
\be
\Gamma^\alpha{}_{\mu\beta}=(\Lambda^{-1})^\alpha{}_\rho\partial_\mu\Lambda^\rho{}_\beta.
\ee
Further imposing the connection to be torsion-free requires the full integrability of the inertial frame so it can be expressed as
\be
\Lambda^\alpha{}_\mu=\partial_\mu\xi^\alpha
\ee
for some functions $\xi^\alpha$. Having removed both the curvature and the torsion, the only remaining geometrical quantity associated to the connection is the non-metricity
\be
Q_{\alpha\mu\nu}\equiv\nabla_\alpha g_{\mu\nu}.
\ee
The form of the connection permits to choose some coordinates where $\cg{\xi}^\alpha=x^\alpha$ and, hence, the connection trivialises $\cg{\Gamma}^\alpha{}_{\mu\nu}=0$. This coordinate system is called the coincident gauge and it is determined up to a global affine transformation $x^\alpha\to A^\alpha{}_\beta x^\beta+c^\beta$. In this gauge, covariant derivatives reduce to partial derivatives so, in particular, the non-metricity becomes
\be
\cg{Q}_{\alpha\mu\nu}=\partial_\alpha g_{\mu\nu}.
\ee
This property is at the heart of the pathological character of theories formulated in the symmetric teleparallel framework \cite{Gomes:2023tur} because an arbitrary action constructed out of the non-metricity $Q_{\alpha\mu\nu}$ and the metric $g_{\mu\nu}$ generically contains a spin-2 particle with derivative interactions which, in the coincident gauge, are not protected by any gauge symmetries. The natural path to obtain sensible theories is to enhance the amount of gauge symmetries in the theory. In this respect, let us notice that the original Diffs invariance in the covariant formulation takes care of the connection degrees of freedom (that are in turn just some Stueckelberg fields \cite{BeltranJimenez:2022azb}). One can extend the Diffs with another copy of Diffs that will then have the task to guarantee that the spin-2 field propagates the two degrees of freedom of GR. This is the Symmetric Teleparallel Equivalent of GR \cite{Nester:1998mp,BeltranJimenez:2017tkd}.

A natural question would be if it is possible to relax the symmetry requirements while still having a non-pathological theory. This is what we intend to explore in this note and we will show how only imposing invariance under TDiffs suffices to have a ghost-free theory.

\section{TDiff Newer GR}
\label{Sec:TDiffNGR}

Let us consider the general quadratic action of symmetric teleparallel theories (sometimes called Newer GR) \cite{BeltranJimenez:2017tkd}
\begin{align}
\mS[g_{\mu\nu},\xi^\alpha]=&\mpl^2\int\diff^4 x\sqrt{-g}\Big[c_1 Q_{\alpha\mu\nu}Q^{\alpha\mu\nu}+c_2 Q_{\alpha\mu\nu}Q^{\mu\nu\alpha}\nonumber\\&+c_3 Q_\alpha Q^\alpha+c_4 \tilde{Q}_\alpha\tilde{Q}^\alpha+c_5 Q_\alpha \tilde{Q}^\alpha\Big]
\label{eq:GenQuad}
\end{align}
with $c_i$ some constant parameters and we have introduced the two non-metricity traces
\be
Q_\alpha=g^{\mu\nu}Q_{\alpha\mu\nu},\qquad \tilde{Q}_\alpha=Q^\mu{}_{\mu\alpha}.
\ee
For a generic choice of parameters, the theory possesses a Diffs symmetry that can be exhausted by going to the coincident gauge. However, the particular choice of parameters 
\be
c_1
= -\frac{c_2}{2}
= - c_3 
= \frac{c_5}{2} = -\frac{1}{8}\,,
c_4= 0\,,
\label{eq:STEGRparams}
\ee
has an enhanced symmetry so that, even in the coincident gauge, there remains a second Diff symmetry (up to a total derivative) that acts trivially on the connection. These are the parameters that correspond to the STEGR, thus describing a massless spin-2 field. 

Out of all the quadratic invariants, let us have a closer look at $Q_\alpha Q^\alpha$. This non-metricity trace (called Weyl trace) can be written as\footnote{The mathematically-minded reader might feel discomfort with the fact that the argument of the logarithm is a negative quantity for Lorentzian metrics. However, all the logarithms that we will encounter will be acted upon by derivatives so there will not be any burden. Either we interpret it as $\nabla_\alpha \log g=\frac{\nabla_\alpha g}{g} $ or we take its principal branch and the derivative will kill the $i\pi$ term.}
\be
Q_\alpha=\nabla_\alpha\log g
\ee
that clearly shows how, in the coincident gauge, the object $Q_\alpha Q^\alpha$ still preserves a TDiff subgroup of Diffs corresponding to Diffs with unitary Jacobian and, hence, it is a promising candidate to deform the STEGR in a sensible way.
If we consider perturbations around Minkowski $g_{\mu\nu}=\eta_{\mu\nu}+h_{\mu\nu}$ in the coincident gauge, we have $\cg{Q}_{\alpha\mu\nu}=\partial_\alpha h_{\mu\nu}$ and it is immediate to see that the quadratic action for the perturbations obtained from \eqref{eq:GenQuad} coincides with the class of theories analysed in \cite{Alvarez:2006uu} for a massless spin-2 field described by $h_{\mu\nu}$. It was shown there that it is necessary to impose a linearised TDiff symmetry to remove ghosts from the spectrum, which amounts to fixing the values \eqref{eq:STEGRparams} but leaving $c_3$ free. Thus, the results in \cite{Alvarez:2006uu} support our claim that deforming the STEGR with the term $Q_\alpha Q^\alpha$ is a promising route where the linearised TDiff is non-linearly completed to full TDiffs. Furthermore, since the linearised TDiff is necessary for having a ghost-free linear spectrum around Minkowski, we could argue that our candidate is in turn unique for a healthy quadratic deformation of STEGR. A caveat of this argument is that the terms $c_2$  and $c_4$ are degenerate in the quadratic action around Minkowski so other extensions based on $\tilde{Q}_\alpha\tilde{Q}^\alpha$ could be envisioned. However, this term does not provide a non-linear completion of the linearised TDiffs (in particular, it does not respect the non-linear TDiff). This means that linearised TDiffs arise as an accidental gauge symmetry of the linear spectrum around Minkowski, thus indicating a discontinuity in the number of degrees of freedom. It is then very likely that the term $\tilde{Q}_\alpha\tilde{Q}^\alpha$ will reintroduce ghosts in the full theory and, certainly, it will lead to strongly coupled modes around Minkowski. This is sufficient for us to discard the $\tilde{Q}_\alpha\tilde{Q}^\alpha$ and, hence, our candidate deformation is indeed unique. Thus, we will focus on the following candidate for a pathology-free extension of the STEGR:
\begin{align}
\mS[g_{\mu\nu},\xi^\alpha]=\frac{\mpl^2}{2}\int\diff^4 x\sqrt{-g}\Big[\Q-\frac{a_5}{4}\; Q_\alpha Q^\alpha\Big]\,,
\label{eq:NGRTDiff0}
\end{align}
with $\Q$ the non-metricity scalar of STEGR, i.e. the quadratic Lagrangian with the parameters \eqref{eq:STEGRparams}, and we parametrise the deviation from STEGR with the constant parameter $a_5$. In the coincident gauge, the action reads
\be
\cg{\mS}[g_{\mu\nu}]=\frac{\mpl^2}{2}\int\diff^4 x\sqrt{-g}\Big[\cg{\Q}-\frac{a_5}{4}\; (\partial\log g)^2\Big],
\label{TDiff}
\ee
where the first term simply describes the STEGR and we have an additional kinetic term for the determinant of the metric. This corresponds to the class of TDiff theories that have been studied in e.g. \cite{Alvarez:2006uu,Alvarez:2009ga,Blas:2011ac,Bello-Morales:2023btf}. The motivation to choose the specific theory \eqref{TDiff} was that it is equivalent to GR in the weak field approximation \cite{Anber:2009qp}. This model has been also analysed in \cite{Pirogov:2011iq} in the class of  unimodular bimode gravity theories. We have seen here that it also arises in a natural way as an extension of GR in the symmetric teleparallel geometry.

Away from the coincident gauge, the Weyl non-metricity trace can be written as
\begin{align}
Q_\alpha=\partial_\alpha\log g-2\Gamma^\mu{}_{\alpha\mu}
=\partial_\alpha\log \frac{g}{\Lambda^2}
\end{align}
where $\Lambda=\det \partial_\mu\xi^\alpha$ and we have used that
\be
\Gamma^\alpha{}_{\mu\alpha}=(\Lambda^{-1})^\alpha{}_\rho\partial_\mu \Lambda^\rho{}_\alpha=\partial_\mu \log \Lambda.
\ee
We can then write the action without fixing any gauge as
\be
\mS[g_{\mu\nu},\xi^\alpha]=\frac{\mpl^2}{2}\int\diff^4 x\sqrt{-g}\left[\Q-\frac{a_5}{4}\; \left(\partial\log \frac{g}{\Lambda^2}\right)^2\right].
\label{eq:TDiffNGR}
\ee
As we have mentioned, this action has one extra scalar degree of freedom in the linear spectrum around Minkowski. Having promoted the TDiff invariance of the linearised theory to the full theory, one could expect to preserve the number of degrees of freedom so the linear spectrum on Minkowski is continuous because the linear TDiff does not arise as an accidental gauge symmetry. Let us notice that this is a common problem of teleparallel theories that are strongly coupled around physically sound backgrounds precisely because of the appearance of accidental gauge symmetries. To show that our TDiff Newer GR has a continuous spectrum, we will integrate out the Stueckelberg fields so that the degrees of freedom are apparent.

\subsection{Integrating out the Stueckelbergs}
\label{Sec:IntStu}

The TDiff Newer GR action written in the form \eqref{eq:TDiffNGR} suggests to trade the combination $g/\Lambda^2$ by a scalar field. Thus, we will introduce a Lagrange multiplier as follows:
\be
\mS=\frac{\mpl^2}{2}\int\diff^4 x\sqrt{-g}\left[\Q-a_5\; (\partial\log \phi)^2-\lambda\left(\phi-\frac{\Lambda}{\sqrt{-g}}\right)\right].
\label{eq:TDiffNGRlambda1}
\ee
The precise form of the constraint imposed by the Lagrange multiplier $\lambda$ has been chosen to decouple the Stueckelberg fields from the metric so they can be integrated out in a straightforward manner. To proceed with the integration of the Stueckelbergs, let us first notice that the determinant of the inertial frame field $\Lambda$ can be written as a divergence:\footnote{We denote the (metric-independent) Levi-Civita symbol by $\epsilon_{\mu\nu\rho\sigma}$.}
\begin{align}
\Lambda=\det\partial_\mu\xi^\alpha &=\frac{1}{4!}\epsilon^{\mu\nu\rho\sigma} \epsilon_{\alpha\beta\gamma\delta}\partial_\mu\xi^\alpha \partial_\nu\xi^\beta \partial_\rho\xi^\gamma \partial_\sigma\xi^\delta\nonumber\\
&=\partial_\mu\Big(\frac{1}{4!}\epsilon^{\mu\nu\rho\sigma} \epsilon_{\alpha\beta\gamma\delta}\xi^\alpha \partial_\nu\xi^\beta \partial_\rho\xi^\gamma \partial_\sigma\xi^\delta\Big)
\end{align}
so, integrating by parts in the last term of the action, that is the only one containing the Stueckelbergs,\footnote{To be precise, the STEGR sector of the action described by $\Q$ also contains the Stueckelberg fields, but they only enter as a total derivative (see e.g. \cite{BeltranJimenez:2019odq}). This implies that the variations of $\Q$ with respect to $\xi^\alpha$ vanish and, hence, such a term will not contribute to the Stueckelberg (or connection) equations of motion. In the rest of the paper we will neglect this total derivative term so, for all our purposes, $\Q$ can be considered to be independent of the Stueckelberg fields or, equivalently, independent of the symmetric teleparallel connection.} we obtain the following action for the Stueckelberg fields:
\begin{align}
\mS_{\text{Stu}}
&\equiv\frac{\mpl^2}{2}\int\diff^4x \lambda \Lambda \nonumber\\
&= -\frac{\mpl^2}{2\times 4!}
\int\diff^4x\partial_\mu\lambda \epsilon^{\mu\nu\rho\sigma} \epsilon_{\alpha\beta\gamma\delta}\xi^\alpha \partial_\nu\xi^\beta \partial_\rho\xi^\gamma \partial_\sigma\xi^\delta.
\label{eq:actionStu}
\end{align}
If we perform variations with respect to the Stueckelbergs, we see that the Levi-Civita symbols prevent the appearance of any second derivatives and the equations of motion are simply
\be
\partial_\mu\lambda\, \epsilon^{\mu\nu\rho\sigma} \epsilon_{\alpha\beta\gamma\delta}\partial_\nu\xi^\beta \partial_\rho\xi^\gamma \partial_\sigma\xi^\delta=0.
\ee
Since the matrix $\partial_\mu\xi^\alpha$ must be non-singular, this equation requires the Lagrange multiplier to be constant $\lambda=\lambda_0$. If we plug this value in the action, the term with the Stueckelberg fields becomes a total derivative and the TDiff Newer GR action reads
\be
\mS=\frac{\mpl^2}{2}\int\diff^4 x\sqrt{-g}\Big[\Q-a_5\; (\partial\log \phi)^2-\lambda_0\phi\Big],
\label{eq:scalarequivalent1}
\ee
that manifestly shows the presence of one single scalar degree of freedom (in addition to the graviton) in the full theory. Furthermore, we immediately see that we need $a_5>0$ for the extra scalar to be healthy. However, we have also obtained a global degree of freedom described by $\lambda_0$ that is reminiscent of integrating out the Stueckelberg fields. This result can be understood in geometrical terms because $\partial_\mu\xi^\alpha$ play the role of integrable frames and the Stueckelberg fields can be associated to the comoving (Lagrangian) coordinates of some medium. In this set-up, $\Lambda\,\diff^4x$ corresponds to the medium volume form and the Stueckelberg action \eqref{eq:actionStu} is nothing but some potential function described by $\lambda$ integrated over the system volume. The medium symmetries force the potential to be constant so that the Stueckelberg fields term is just a {\it cosmological constant} for the medium. More explicitly, the Stueckelberg fields action can be recast as an integration over the Lagrangian coordinates as
\be
\mS_{\text{Stu}}=\int\diff^4x \lambda(x)\det\partial_\mu\xi^\alpha=\int\diff^4\,\xi \lambda(x(\xi)),
\ee
whose variation trivially imposes the constancy of $\lambda$ (let us notice that the fields in the last form of the action are the $x$'s). 

The interesting feature of the scalar field equivalent \eqref{eq:scalarequivalent1} is that the global degree of freedom $\lambda_0$ is coupled to the scalar in its potential. In fact, the sector with a trivial $\lambda_0$ also trivialises the scalar field potential. We will come back to this point later, but first let us discuss another procedure to obtain this result by resorting to an equivalent way of introducing Stueckelberg fields that has been used in the literature.

\subsection{An alternative Stueckelbergisation}

The fact that Diffs can be restored in different ways within the framework of symmetric teleparallel geometries has already been discussed in \cite{BeltranJimenez:2022azb}. We will here use yet another Stueckelbergisation \`a la Henneaux-Teitelboim \cite{Henneaux:1989zc} originally introduced in the context of unimodular gravity and applied in the context of TDiff scalar theories in \cite{Jaramillo-Garrido:2024tdv}. We start from the action in the coincident gauge \eqref{TDiff} and restore Diffs by introducing the scalar density $\partial_\mu(\sqrt{-g}A^\mu)$ such that $\frac{1}{\sqrt{-g}}\partial_\mu(\sqrt{-g}A^\mu) = D_\mu A^\mu$, with $D_\mu$ the covariant derivative associated to the Levi-Civita connection of the metric, transforms as a true scalar. The coincident gauge is then recovered by choosing the coordinates where $\partial_\mu(\sqrt{-g}A^\mu)$ is constant, i.e., $D_\mu A^\mu\propto\sqrt{-g}$. By using the explained trick, we can write the equivalent action
\be
\cg{\mS}=\frac{\mpl^2}{2}\int\diff^4 x\sqrt{-g}\Big[\cg{\Q}- \frac{a_5}{(D\cdot A)^2}\,(\partial D\cdot A)^2\Big].
\ee
We can corroborate that the action \eqref{TDiff} is recovered in the coincident gauge where $D_\mu A^\mu\propto\sqrt{-g}$ and we still have the characteristic invariance under global affine transformations of the coincident gauge explained above. In this formulation, we have introduced an additional gauge symmetry given by $A^\mu\to A^\mu+D_\nu \theta^{[\mu\nu]}$ for an arbitrary $\theta^{\mu\nu}$ or, equivalently, $A^\mu\to A^\mu+\vartheta^\mu$ with $D_\mu \vartheta^\mu = 0$. Let us notice that this symmetry is irrelevant for the coincident gauge that corresponds to a choice of coordinates. It is not difficult to see that the one-form field $A_\mu$ is proportional to the Hodge dual of the three form $\epsilon_{\alpha\beta\gamma\delta} \xi^\alpha\diff\xi^\beta\wedge\diff\xi^\gamma\wedge\diff\xi^\delta$, i.e.,
\be
\bf{A}\propto\epsilon_{\alpha\beta\gamma\delta}\xi^\alpha\star\big(\diff\xi^\beta\wedge\diff\xi^\gamma\wedge\diff\xi^\delta\big)
\ee
so the Henneaux-Teitelboim restoration of Diffs is dual to the more direct Stueckelbergisation of the previous section.\footnote{The Stueckelberg fields $\xi^\alpha$ that originate within the symmetric teleparallel framework are analogous to the ones employed in \cite{Kuchar:1991xd} in the context of unimodular gravity and its dual relation with the Henneaux-Teitelboim procedure was also noticed in that work.} The restoring vector field $A^\mu$ can be integrated out following a procedure similar to the integration of the $\xi$'s. We first rewrite the action as
\be
{\mS}=\frac{\mpl^2}{2}\int\diff^4 x\sqrt{-g}\Big[\Q- a_5\,(\partial \log\phi)^2-\lambda\big(\phi- D\cdot A\big)\Big].
\label{eq:Stuelambda2}
\ee
The equation for $A^\mu$ imposes $\partial_\mu\lambda=0$ which means that the Lagrange multiplier is constant $\lambda=\lambda_0$. We then insert this constraint in the action to obtain the equivalent action
\be
\mS=\frac{\mpl^2}{2}\int\diff^4 x\sqrt{-g}\Big[\Q-a_5\,(\partial \log \phi)^2-\lambda_0\phi\Big],
\label{eq:Actionlambda1}
\ee
which coincides with \eqref{eq:scalarequivalent1}, as it should. This proves the equivalence of the two Stueckelbergisation procedures and confirms that the TDiff teleparallel theory under consideration contains one additional scalar local degree of freedom.

\subsection{The scalar field equivalent}

We have shown in two alternative ways that the TDiff Newer GR theory can be written as the scalar field theory \eqref{eq:Actionlambda1}. This theory can be expressed in a more canonical form by redefining $\phi= e^{\frac{\varphi}{\sqrt{a_5}\mpl}}$ so we have
\be
\mS=\int\diff^4 x\sqrt{-g}\left[\frac{\mpl^2}{2}\Q - \frac{1}{2}\,(\partial \varphi)^2-\frac{\mpl^2\lambda_0}{2}e^{\frac{\varphi}{\sqrt{a_5}\mpl}}\right].
\label{Eq:scalarequivalent}
\ee
Thus, the TDiff Newer GR describes one additional scalar field with an exponential potential plus a global degree of freedom given by $\lambda_0$. This result agrees with previous findings in the literature (see e.g. \cite{Pitts:2001jw,Alvarez:2006uu,Lopez-Villarejo:2010uib,Blas:2011ac}) where TDiff theories have been explored from different perspectives and their equivalence to scalar field theories plus an integration constant has also been found. Expressed in the equivalent form \eqref{Eq:scalarequivalent} it is apparent that the theories \eqref{eq:NGRTDiff0} recover GR in the weak field limit in the sense that they will reproduce the same PPN parameters. This was the main motivation for their consideration in \cite{Bello-Morales:2023btf} and might be slightly more obscure in the formulation \eqref{TDiff}. It might be worth emphasising that the additional scalar degree of freedom that we have unveiled does not feature any coupling to matter derived from our procedure of integrating out the Stueckelberg fields. Thus, unless the matter sector also features TDiff couplings (see Sec. \ref{Sec:MC} below) from the onset, this sector is oblivious to the extra scalar, which is crucial for recovering the GR PPN parameters. In this sense, these theories are clearly different from e.g. Brans-Dicke, $f(R)$ or $f(G)$, with $R$ and $G$ the metric Ricci scalar and Gauss-Bonnet respectively, where there is an extra scalar field which does couple to matter in the Einstein frame). Among other consequences, this means that the quadratically deformed STEGR described by \eqref{eq:NGRTDiff0} is less prone to be immediately ruled out by local gravity tests.

It is important to emphasise that $\lambda_0$ is not a parameter of the theory, but a global degree of freedom to be fixed by boundary conditions. In this respect, the choice of $\lambda_0$ could be interpreted as a selection rule of the theory. Another remarkable property of the resulting theory is that the generated interactions from the integration of the Stueckelbergs are local. This is not usually the case and integrating out Stueckelberg fields commonly give rise to non-local operators. A paradigmatic example is Proca theory for a massive spin-1 field which, after integrating out the Stueckelberg fields, is described by the Lagrangian $\mathcal{L}_{\text{Proca}}=-\frac{1}{4}F_{\mu\nu}(1-\frac{m^2}{\Box})F^{\mu\nu}$ that is non-local.\footnote{A similar result arises for massive gravity and the non-local operators can be interpreted as filters that could be used for degravitating \cite{Dvali:2007kt}.}

Let us finally comment on the issue of inserting a solution like $\lambda=\lambda_0$ in the action that could cause some doubts. Although this is a legitimate procedure (provided it is appropriately done), one could worry that some extra care should be taken for instance to maintain the variational principle. In our case, there is not any particularly subtle point, but it is however reassuring to show explicitly that nothing goes wrong by checking the equivalence of the equations of motion. This can be checked with both Stueckelbergisation procedures that we have discussed, but for the sake of concreteness we will do it for \eqref{eq:Stuelambda2}. We can integrate by parts in that action to write it as
\be
\mS=\frac{\mpl^2}{2}\int\diff^4 x\sqrt{-g}\Big[\Q- a_5\,(\partial \log\phi)^2-\lambda \phi-A^\mu\partial_\mu\lambda\Big].
\ee
It is obvious that $\lambda$ will enter as a source in the scalar field equation and will contribute a term $\propto \lambda \phi g_{\mu\nu}$ to the energy-momentum tensor of $\phi$. The Lagrange multiplier will contribute another term to the gravitational equations given by $g_{\mu\nu}A^\alpha\partial_\alpha\lambda $. On the other hand, the equation for $A^\mu$ still imposes $\lambda$ to be constant which can be used in the scalar and gravitational equations. We then find that the term $(A^\alpha\partial_\alpha\lambda) g_{\mu\nu}$ in the gravitational equations vanishes while the other contributions are exactly the same as those derived from \eqref{eq:Actionlambda1}, thus showing the validity of our procedure.

\subsection{Symmetries}

After discussing the equivalence of the TDiff Newer GR theory to a scalar field theory with a global degree of freedom, a discussion on the symmetries of the theory is in order. Let us start by the most trivial one, i.e., the Diffs shared by all the symmetric teleparallel theories away from the coincident gauge. They correspond to the usual Diffs with the Stueckelbergs changing as
\be
\xi^\alpha(x)\to \tilde{\xi}^\alpha(\tilde{x})=\xi^\alpha(x)
\ee
so we have
\be
\Lambda^\alpha{}_\mu(x)\to\tilde{\Lambda}^\alpha{}_\mu(\tilde{x})=\frac{\partial x^\nu}{\partial \tilde{x}^\mu}\Lambda^\alpha{}_\nu(x).
\ee
This symmetry will be responsible for the usual Bianchi identities relating the metric and the connection field equations. The TDiff Newer GR takes its name from the additional symmetry it realises in the coincident gauge that is apparent from \eqref{TDiff}. It is clear that, in the coincident gauge, we can still perform a TDiff that leaves the Stueckelberg fields untouched, i.e., the metric changes but the connection does not. Of course, the symmetry remains away from the coincident gauge and the simplest realisation is the same as in the coincident gauge, i.e., a TDiff coordinate transformation $x^\mu\to\tilde{x}^\mu(x)$ with $\det \frac{\partial \tilde{x}^\mu}{\partial x^\nu}=1$ but leaving the Stueckelberg fields unchanged
\be
\xi^\alpha(x)\to \xi^\alpha(\tilde{x}).
\label{eq:StuckTDiff}
\ee
This means that the inertial frame transforms as a scalar:
\be
\Lambda^\alpha{}_\mu(x)\to\tilde{\Lambda}^\alpha{}_\mu(\tilde{x})=\frac{\partial\xi^\alpha(\tilde{x})}{\partial \tilde{x}^\mu}=\frac{\partial\xi^\alpha(x)}{\partial x^\mu}=\Lambda^\alpha{}_\mu(x).
\ee
In order to distinguish the usual Diffs (under which the reference frame transforms as a 1-form) from these Diffs (under which the reference frame transforms as a scalar), we will refer to the latter as Diffs'.

There is another realisation of the TDiff symmetry that is realised entirely on the Stueckelberg fields. Since they only enter through the determinant of the inertial frame matrix $\Lambda^\alpha{}_\mu$, we can perform an arbitrary field redefinition of the Stueckelberg fields
\be
\xi^\alpha\to\xi^\alpha(\zeta)\qquad\text{with}\qquad\left\vert\det\frac{\partial\xi^\alpha}{\partial\zeta^\beta}\right\vert=1,
\ee
which corresponds to an internal TDiff transformation. This realisation is practical because it  does not involve spacetime transformations. The two discussed realisations are the limiting cases of the general realisation that involves both Diffs' and field redefinitions of the following form:
\be
\xi^\alpha\to\xi^\alpha(\zeta(\tilde{x})).
\ee
Under this transformation, the inertial frame changes as
\begin{align}
\tilde{\Lambda}^\alpha{}_\mu(\tilde{x})=\frac{\partial\xi^\alpha(\zeta(\tilde{x}))}{\partial \tilde{x}^\mu}=\frac{\partial\xi^\alpha}{\partial \zeta^\beta}\frac{\partial\zeta^\beta(x)}{\partial x^\mu}
\end{align}
so we find the transformation rule
\be
\frac{g}{\Lambda^2}\to \left(\det \frac{\partial \tilde{x}^\mu}{\partial x^\nu}\det\frac{\partial\xi^\alpha}{\partial \zeta^\beta}\right)^{-2}\frac{g}{\Lambda^2}.
\ee
From this expression, we see that we need to impose
\be
\left\vert\det \frac{\partial \tilde{x}^\mu}{\partial x^\nu}\det\frac{\partial\xi^\alpha}{\partial \zeta^\beta}\right\vert=1
\label{eq:generalTDifftransf}
\ee
to have invariance. From this relation we corroborate that the two TDiff realisations discussed above are recovered by trivialising the spacetime Diffs' and the field redefinitions respectively. In fact, these are the two extreme cases where a proper TDiff arises. In general, we can perform arbitrary Diffs' and field redefinitions as long as the non-TDiff pieces compensate and the constraint \eqref{eq:generalTDifftransf} is satisfied.

The symmetries that we have discussed so far will be symmetries of any theory that depends on the combination $g/\Lambda^2$. The symmetric teleparallel theories however have more symmetries because the fundamental quantity is $Q_\alpha=\partial_\alpha\log\frac{g}{\Lambda^2}$. This means that we can perform an arbitrary transformation that violates the condition \eqref{eq:generalTDifftransf} as long as it is global. In that case, both Jacobians $\det \frac{\partial \tilde{x}^\mu}{\partial x^\nu}$ and $\det\frac{\partial\xi^\alpha}{\partial \zeta^\beta}$ are constant so they are killed by the derivative in $Q_\alpha$. Thus, these global transformations will be symmetries of the symmetric teleparallel theories. We should notice that this global symmetry  has only one parameter that is associated to the product of the two Jacobians. In other words, an arbitrary global transformation can always be factorised as a dilation (i.e., a transformation proportional to the identity) and a transformation subject to \eqref{eq:generalTDifftransf} so only the global dilation piece is actually an additional global symmetry. More explicitly, this global symmetry can be entirely realised on the Stueckelberg fields as $\xi^\alpha\to\kappa\xi^\alpha$, with $\kappa$ constant or, in the coincident gauge, as a usual dilation $x^\mu\to\kappa x^\mu$. How does this additional symmetry presents itself in the scalar equivalent? A priori, this global dilation should correspond to a shift symmetry in \eqref{Eq:scalarequivalent}, but it would seem that the exponential potential breaks the would-be shift symmetry. However, we must note that $\lambda_0$ is not an external parameter, but a global degree of freedom. Thus, if we perform a shift $\varphi\to \varphi+c$ in \eqref{eq:Actionlambda1} together with a rescaling $\lambda_0\to \lambda_0e^{-\frac{c}{\sqrt{a_5}\mpl}}$, the action remains invariant. The consequence of this symmetry is the obvious relation $\frac{\partial \mathcal{L}(\phi)}{\partial \varphi}=\frac{\lambda_0}{\sqrt{a_5}\mpl}\frac{\partial \mathcal{L}(\phi)}{\partial \lambda_0}$. More interesting is the on-shell conserved current that is obtained for the formulation in terms of the Stueckelberg fields, which can be written as
\be
J^\mu=\sqrt{-g}Q^\mu-\frac14\partial_\alpha\big(\sqrt{-g} Q^\alpha\big)(\Lambda^{-1})^\mu{}_\beta \xi^\beta.
\label{eq:conservedJ}
\ee
Since the Stueckelberg equations can be written as 
\be
\partial_\mu\Big[\partial_\alpha\big(\sqrt{-g} Q^\alpha\big)(\Lambda^{-1})^\mu{}_\beta\Big]=0,
\ee
we see that $\partial_\mu J^\mu=0$ upon use of the above equations so it is indeed conserved on-shell.

In general, the symmetric teleparallel theories have the following set of off-shell conserved currents:
\be
J^\mu{}_\alpha=\epsilon^{\mu\nu\rho\sigma} \epsilon_{\alpha\beta\gamma\delta}\partial_\nu\xi^\beta \partial_\rho\xi^\gamma \partial_\sigma\xi^\delta.
\ee
These off-shell currents are nothing but the derivatives of the determinant of the inertial frame $J^\mu{}_\alpha=\frac{\partial\Lambda}{\partial\Lambda^\alpha{}_\mu}$ so the Stueckelberg equations can also be written as
\be
\partial_\mu\left(\frac{\delta \mathcal{S}}{\delta \Lambda} J^\mu{}_\alpha\right)=0.
\ee
Since $J^\mu{}_\alpha$ are conserved off-shell, this equation implies
\be
\frac{\delta \mathcal{S}}{\delta \Lambda}=\lambda_0
\label{eq:genlambda0}
\ee
for some constant $\lambda_0$. This shows how the global degree of freedom emerges in the original formulation of the theory \eqref{eq:TDiffNGR}. Furthermore, it represents a generalisation of the result shown in \ref{Sec:IntStu} of the appearance of a global degree of freedom because here we have not specified any particular action to obtain \eqref{eq:genlambda0} , thus showing that the appearance of a global degree of freedom is in fact a generic property of TDiff symmetric teleparallel theories, as already noticed in e.g. \cite{Pitts:2001jw,Alvarez:2006uu,Lopez-Villarejo:2010uib,Blas:2011ac}.

\subsection{Cosmology}
\label{Sec:Cosmology}

In the cosmological context, the behaviour of this type of scalar field models with 
an exponential potential \eqref{Eq:scalarequivalent} is well-known \cite{Copeland:1997et}. Interestingly, the cosmologies based on \eqref{Eq:scalarequivalent} precisely correspond to the models whose integrability can be related to the existence of an additional conserved quantity \cite{Chiba:2024iia}.  We will provide here the main features for these cosmologies in the presence of an additional matter sector $\mathcal{S}_F=\int\diff^4x\sqrt{-g}\mathcal{L}_F$ with $\mathcal{L}_F$ the Lagrangian of a barotropic fluid with energy density $\rho_F$ and constant equation of state $-1 < \omega_F < 1$. The cosmological equations for a Friedmann-Lema\^itre-Robertson-Walker universe described by the line element
\be
\diff s^2=-\diff t^2+a^2(t)\diff\vec{x}^2,
\ee
with $a(t)$ the scale factor, are the usual Friedmann equations
\be
\left(\frac{\dot{a}}{a}\right)^2=\frac{8\pi G}{3}\rho,\quad\frac{\ddot{a}}{a}=-\frac{4\pi G}{3}(\rho+3p)
\ee
with 
\begin{eqnarray}
\rho&=&\frac12\dot{\varphi}^2+\frac{\mpl^2\lambda_p}{2}e^\frac{\varphi}{\sqrt{a_5}\mpl}+\rho_F,\\
p&=&\frac12\dot{\varphi}^2-\frac{\mpl^2\lambda_p}{2}e^\frac{\varphi}{\sqrt{a_5}\mpl}+\omega_F\rho_F.
\end{eqnarray}
In these scenarios, there exist two types of late-time attractors for $\lambda_0 > 0$: 
 \begin{itemize}
 \item For $1/a_5<3(1+\omega_F)$ and $a_5 > 1/6$, 
the scalar field dominates at late times with an effective equation of state $\omega_S=-1+1/(3a_5)$
\item For $1/a_5>3(1+\omega_F)$ we have scaling solutions in which the scalar field energy density mimics the barotropic fluid with $\Omega_S=3a_5(1+\omega_F)$. Constraints from light element abundances
on the expansion rate of the universe during nucleosynthesis allows to set a bound on the $a_5$ parameter $a_5 \lsim 1/20$ in this case.
\end{itemize}

These results are in agreement with what was obtained in \cite{Bello-Morales:2023btf} for the TDiff model in \eqref{TDiff}. In addition, it has been shown in  \cite{Bello-Morales:2023btf} that in case $\varphi = \text{cte}$, $\lambda_0 = 0$, which corresponds to $\partial g = 0$, we recover GR cosmology for a barotropic fluid, although that solution is unstable for all values of $\omega_F > -1$ \cite{Copeland:1997et}. When the scalar field is subdominant with respect to the barotropic fluid, it behaves as a cosmological constant throughout the cosmic history until it reaches the late-time attractor behaviour \cite{Bello-Morales:2023btf}.

On the other hand, for negative $\lambda_0$, which corresponds to  
$\omega_S > 1$, we have recollapsing solutions for  $a_5 \geq 1/6$ and in the case $a_5 < 1/6$, we have 
ever-expanding solutions for $\rho_F = 0$  (see \cite{Bello-Morales:2023btf}) which corresponds to a kinetic-dominated attractor in \cite{Heard:2002dr}. However, these expanding solutions with a negative potential, apart from introducing potential tachyon instabilities,  are also unstable in the presence of ordinary matter, spatial curvature or anisotropic shear and always recollapse to a singularity \cite{Heard:2002dr}.

Let us notice that the branch with $\lambda_0=0$ can accommodate homogeneous but anisotropic solutions with $\varphi=\phi_0(t)+v_i x^i$, with $v_i$ a constant vector. The reason is that the shift symmetry for the field $\varphi\to\varphi+ c$ is exact in the sense that it does not involve $\lambda_0$. This shift symmetry then allows to restore homogeneity. The realisation of homogeneity corresponds to a spatial translation $\vec{x}\to\vec{x}+\vec{x}_0$ in combination with an internal shift with parameter $c=-\vec{v}\cdot\vec{x}_0$. The cosmological scenarios for these configurations correspond to having a homogeneous non-comoving fluid and they could provide a realisation of the scenarios explored in \cite{Maroto:2005kc,BeltranJimenez:2007rsj,BeltranJimenez:2008rei,Cembranos:2019plq} (see also \cite{Garcia-Garcia:2016dcw} for a bi-metric realisation).

In terms of the scalar equivalent, it is clear that the cosmological solutions for homogeneous and isotropic universes will be supported by a homogeneous scalar field. However, from the perspective of the Stueckelberg fields the situation is less simple in view of the classification performed in \cite{Gomes:2023hyk} for the inequivalent cosmological branches of teleparallel theories. Based on this classification, the different non-trivial realisations of homogeneity and isotropy was adapted to the symmetric teleparallel framework in \cite{Gomes:2023tur} for the spatially flat cosmologies, where it was also explained how the cosmological symmetries are non-trivially realised on the Stueckelberg fields. There are three different cosmologies that correspond to the following Stueckelberg field configurations \cite{Gomes:2023tur}:
\begin{itemize}
    \item Trivial branch: 
    \begin{equation}
    \xi^0=\xi(t),\quad\xi^i=\sigma_0x^i.\label{eq:trivialbranch}
    \end{equation} 
    \item Non-trivial branch I:
    \begin{equation}
        \xi^0=\xi(t)-\frac12\sigma_0\omega|\vec{x}|^2,\quad\xi^i=\sigma_0x^i.
    \end{equation}
    \item Non-trivial branch II:
    \begin{equation}
    \xi^0=\xi(t),\quad\xi^i=\Big[\omega\xi(t)+\sigma_0\Big] x^i.
    \end{equation}
\end{itemize}
In these expressions, $\sigma_0$ and $\omega$ are constant parameters and $\xi(t)$ is an arbitrary function of time (that describes the single cosmological degree of freedom of the background connection). As explained in \cite{Gomes:2023tur}, the general symmetric teleparallel theories feature a global symmetry $\xi^\alpha\to A^\alpha{}_\beta\xi^\beta+b^\alpha$ that can be used to compensate for spatial rotations and translations, that do not leave the above configurations invariant by themselves. However, the TDiff-invariant case has an enhanced symmetry and now we can perform an arbitrary field redefinition satisfying \eqref{eq:StuckTDiff}. It turns out that the non-trivial branches can be related with the trivial branch by exploiting this enhanced symmetry. It is instructive to see explicitly how this comes about.
\begin{itemize}
    \item Non-trivial branch I. For this configuration, we can perform the TDiff field redefinition:
    \be
\zeta^0=\xi^0+\frac{\omega}{2\sigma_0}\vec{\xi}^2,\quad \zeta^i=\xi^i
    \ee
so we recover the trivial branch with $\zeta^0=\xi(t)$ and $\zeta^i=\sigma_0x^i$. The Jacobian of the transformation is easy to compute. Since $\frac{\partial\zeta^i}{\partial \xi^0}=0$, the Jacobian is simply $\det\frac{\partial\zeta^\alpha}{\partial \xi^\beta}=\frac{\partial\zeta^0}{\partial \xi^0}\det \frac{\partial\zeta^i}{\partial \xi^j}=1$ so the field redefinition indeed satisfies the TDiff condition.

\item Non-trivial branch II. This branch can be brought to the trivial branch by performing the field redefinition
\be
\zeta^0=\frac{(\omega\xi^0+\sigma_0)^4}{4\omega\sigma_0^3},\quad \zeta^i=\frac{\sigma_0}{\omega\xi^0+\sigma_0}\xi^i.
\ee
In this case we must identify $\frac{(\omega\xi^0+\sigma_0)^4}{4\omega\sigma_0^3}$ with the function $\xi(t)$ in \eqref{eq:trivialbranch}. The Jacobian of this transformation can be computed as in the previous case since $\frac{\partial\zeta^0}{\partial \xi^i}=0$ so that we obtain 
\begin{eqnarray}
\det\frac{\partial\zeta^\alpha}{\partial \xi^\beta}=\frac{\partial\zeta^0}{\partial \xi^0}\det \frac{\partial\zeta^i}{\partial \xi^j}=\frac{\partial\zeta^0}{\partial \xi^0}\left(\frac{\sigma_0}{\omega\xi^0+\sigma_0}\right)^3=1.
\end{eqnarray}
\end{itemize}
Thus, we confirm that the three cosmological branches are related via a TDiff field redefinition of the Stueckelberg fields and, hence, they are physically equivalent configurations that pleasingly recover the single branch of the scalar equivalent with a time-dependent profile. Furthermore, we can trivially check the constancy of the Lagrange multiplier in the cosmological scenario by noticing that $\Lambda=\sigma_ 0^3\dot{\xi}$, as it should because we have shown that $\Lambda$ is always a total derivative, so its equation of motion yields $\dot{\lambda}=0$, i.e., $\lambda$ is constant. This is the cosmological minisuperspace realisation of the general result obtained in \eqref{eq:genlambda0}.

\section{Beyond TDiff Newer GR}
\label{Sec:GFExt}

In the precedent section we have discussed how to extend the quadratic Newer GR action of the symmetric teleparallel framework in a ghost-free way by relaxing the full Diff symmetry of STEGR to its TDiff subgroup. This procedure is not unique of the quadratic theory and in turn it can be straightforwardly extended in various directions. In this section, we will present the non-linear extensions to include an arbitrary function of $Q_\alpha Q^\alpha$ that will result in a specific type of $K-$essence for the scalar equivalent, and an extension including derivatives of $Q_\alpha$ that will be related to shift-symmetric Horndeski theories.

\subsection{Symmetric teleparallel $K$-essence}
Clearly, the STEGR can be extended to a more general theory where we add an arbitrary function of the non-metricity trace
\be
\mS=\int\diff^4 x\sqrt{-g}\Big[\frac{\mpl^2}{2}\Q+\mathcal{K}(Q_\alpha Q^\alpha)\Big]
\label{eq:SKessence}
\ee
so, {\it mutatis mutandis}, the procedure employed for the TDiff Newer GR will lead to an equivalent family of $K$-essence theories with the same linear potential determined by the integration constant $\lambda_0$. Thus, we can write the theory described by \eqref{eq:SKessence} in the following equivalent form:
\be
\mS=\int\diff^4 x\sqrt{-g}\left[\frac{\mpl^2}{2}\Q+\mathcal{K}(X)-\lambda_0e^{\varphi}\right]\, ,
\label{eq:SymKessence}
\ee
with $X\equiv\partial_\alpha \varphi \partial^\alpha \varphi$. The appearance of the integration constant can be obtained directly from the original action \eqref{eq:SKessence}, noting that the equation for the Stueckelbergs can be written as
\be
\partial_\mu\left(\frac{\delta\mS}{\delta \Lambda}\Lambda (\Lambda^{-1})^\mu{}_\alpha\right)=0.
\ee
If we use now that
\be
\Lambda (\Lambda^{-1})^\mu{}_\alpha\propto \epsilon^{\mu\nu\rho\sigma}\epsilon_{\alpha\beta\gamma\delta} \Lambda^\beta{}_\nu \Lambda^\gamma{}_\rho \Lambda^\delta{}_\sigma
\ee
together with the property $\partial_{[\mu} \Lambda^\alpha{}_{\nu]}=0$, the Stueckelberg fields equation reduces to 
\be
\partial_\mu\left(\frac{\delta\mS}{\delta \Lambda}\right)=0,
\ee
that can be integrated to obtain
\be
\frac{\delta\mS}{\delta \Lambda}=\lambda_0.
\ee
This is the generalisation of the result obtained in Sec. \ref{Sec:IntStu} where $\frac{2}{\mpl^2}\frac{\delta\mS}{\delta \Lambda}=\lambda$ for the formulation \eqref{eq:TDiffNGRlambda1} for instance and it is an explicit example of the general result shown in \eqref{eq:genlambda0}. It is evident that all the discussions in the precedent sections regarding cosmological branches, symmetries, etc. will also be valid for the $K$-essence family. In particular, the global dilation invariance remains and the corresponding conserved current derived from \eqref{eq:SKessence} is given by
\be
J^\mu=\sqrt{-g}\mathcal{K}'Q^\mu-\frac14\partial_\alpha\Big(\sqrt{-g}\mathcal{K}' Q^\alpha\Big)(\Lambda^{-1})^\mu{}_\beta \xi^\beta,
\label{eq:conservedJK}
\ee
which is conserved upon use of the connection equation
\be
\partial_\mu\Big[\partial_\alpha\big(\sqrt{-g}\mathcal{K}' Q^\alpha\big)(\Lambda^{-1})^\mu{}_\beta\Big]=0.
\ee
This dilation invariance translates into a shift symmetry together with a re-scaling of the global degree of freedom $\lambda_0$, which is apparent in \eqref{eq:SymKessence}. In this respect, since a superfluid can be described in terms of a shift symmetric scalar field whose leading order interactions are given by a $\mathcal{K}(X)$ theory, it is tempting to interpret \eqref{eq:SymKessence} as a special type of superfluid. For the solutions with a trivial global degree of freedom $\lambda_0=0$, the interpretation as a superfluid is immediate and, as a matter of fact, the homogeneous and anisotropic configurations discussed in Sec. \ref{Sec:Cosmology} can be nicely interpreted as a moving superfluid. On the other hand, for non-trivial values of $\lambda_0$, we could interpret it as a superfluid coupled to an external global degree of freedom that preserves the super-fluid symmetries. A more thorough exploration of this interpretation  would be interesting to pursue further.

\subsection{TDiff symmetric teleparallel Horndeski}
\label{Sec:TDiffHorndeski}

In the previous TDiff symmetric teleparallel $K$-essence theories, we have restricted ourselves to ultralocal interactions of $Q_\alpha$. It is however possible to also include interactions featuring derivatives of the non-metricity Weyl trace, although this must be done with some care. The reason is that, adding more derivatives, we will be prone to introducing new (ghostly) degrees of freedom. However, guided by the Horndeski scalar-tensor theories \cite{Horndeski:1974wa} (see also \cite{Deffayet:2009wt,Deffayet:2011gz}), we just need to consider the permitted derivative non-minimal couplings of a scalar field and translate to our framework. As an illustrative example of the procedure, let us consider the quadratic TDiff symmetric teleparallel action supplemented with a {\it non-minimal coupling} to the metric Einstein tensor:
\be
\mS=\frac{\mpl^2}{2}\int\diff^4x\sqrt{-g}\left[\Q-\frac{a_5}{4}Q_\alpha Q^\alpha+
\frac{\beta}{2} G^{\mu\nu}Q_\mu Q_\nu\right].
\label{eq:HorndeskTDiff}
\ee
One might object that we are mixing the symmetric teleparallel and the usual GR frameworks. However, one could also argue that we are not abandoning the symmetric teleparallel arena because the Einstein tensor can be covariantly expressed in terms of the non-metricity, as can be easily proven from the relation
\be
G_{\mu\nu}(g)=\frac{1}{\sqrt{-g}}\frac{\delta (\sqrt{-g} \Q)}{\delta g^{\mu\nu}}.
\ee
In fact, we could use this relation to rewrite the action in the suggestive form
\begin{align}
\mS=\frac{\mpl^2}{2}\int\diff^4x\Big[&\sqrt{-g}\Q+
\frac{\beta}{2} \frac{\delta (\sqrt{-g} \Q)}{\delta g^{\mu\nu}}Q^\mu Q^\nu\nonumber\\
&-\frac{a_5}{4}\sqrt{-g} Q_\alpha Q^\alpha
\Big].
\end{align}
that is manifestly written in the symmetric teleparallel realm. In any case, the point that we want to stress is that, after the procedure of introducing the Stueckelberg fields as
\begin{align}
\mS=&\frac{\mpl^2}{2}\int\diff^4 x\sqrt{-g}\Big[\Q-a_5\; (\partial\log \phi)^2\nonumber\\&+
2\beta G^{\mu\nu}\partial_\mu\log\phi \partial_\nu\log\phi
-\lambda\left(\phi-\frac{\Lambda}{\sqrt{-g}}\right)\Big]
\end{align}
and integrate them out as in the previous sections so we obtain $\lambda=\lambda_0$, it should be clear that we will simply generate a non-minimal derivative coupling of the scalar field to the Einstein tensor. The scalar equivalent will then be
\begin{align}
\mS=\int\diff^4x\sqrt{-g}\Bigg[&\frac{\mpl^2}{2}\Q+\beta G^{\mu\nu}\partial_\mu \varphi \partial_\nu\varphi\nonumber\\
&-\frac{1}{2}(\partial\varphi)^2-\frac{\mpl^2\lambda_0}{2}e^{\frac{\varphi}{\sqrt{a_5}\mpl}}\Bigg],
\end{align}
where we have canonically normalised the scalar field $\phi=e^{\frac{\varphi}{\sqrt{a_5}\mpl}}$ and we have redefined $\beta\to \beta/a_5$. This action is well-known to be ghost-free, since it belongs to the Horndeski class. Obviously, we could have considered a generic $K$-essence term instead of the quadratic term described by $a_5$. In fact, the depicted procedure can be straightforwardly extended to obtain the whole class of shift symmetric Horndeski Lagrangians with an exponential potential that couples to the global degree of freedom. The theory described by \eqref{eq:HorndeskTDiff} is a particular case of the quartic Horndeski Lagrangian that would be straightforward to extend to the general quartic Horndeski case. It is interesting and instructive to show how the cubic Horndeski Lagrangian can be formulated as a specific TDiff symmetric teleparallel theory. Let us consider an action that includes the following term:
\be
\mS_{\text{cubic}}=\frac12\int\diff^4x\sqrt{-g}Q^\mu \nabla_\mu G_{3}(Q_\alpha Q^\alpha),
\label{eq:CubicH}
\ee
where $G_3$ is an arbitrary function and we assume that the complete action can contain other sectors constructed in the TDiff symmetric framework as discussed above, e.g., a $K$-essence sector and/or the non-minimal coupling introduced in \eqref{eq:HorndeskTDiff}. Using once again the same procedure to integrate the Stueckelbergs out, the action \eqref{eq:CubicH} can be written as
\begin{equation}
\mS_{\text{cubic}}=\int\diff^4x\sqrt{-g}\Big[\partial^\mu\log\phi \partial_\mu G_{3}\big((\partial\log\phi)^2\big)-\lambda_0\phi\Big].
\end{equation}
We can now integrate by parts and redefine the scalar field as $\varphi=\log\phi$ to finally obtain
\be
\mS_{\text{cubic}}=-\int\diff^4x\sqrt{-g}\Big[G_{3}\big(X\big){\Box}\varphi +\lambda_0e^{\varphi}\Big],
\ee
with ${\Box}\equiv D_\mu D^\mu$. We thus see the advertised equivalence to the cubic Horndeski. In particular, let us notice that we recover a cubic Galileon \cite{Nicolis:2008in} for $G_3\propto Q^2$ and a trivial global degree of freedom $\lambda_0=0$. The Galileon is special because it enjoys an additional Galilean symmetry $\varphi\to \varphi+c+v_\mu x^\mu$. In the TDiff symmetric teleparallel formulation, this symmetry only exists in the sector of trivial $\lambda_0$. Interestingly,  while the pure cubic Galileon (i.e., $\lambda_0=0$) has been shown to conflict with cosmological observations \cite{Renk:2017rzu,Peirone:2017vcq}, its generalisation with a potential has been claimed to be observationally favoured in \cite{Ye:2024kus}. As we have demonstrated, the generation of such a potential for the cubic Galileon appears very naturally in the TDiff symmetric teleparallel framework. It is also pertinent to mention that the sector with $\lambda_0=0$ can also be made viable provided \eqref{eq:CubicH} is supplemented with a $K$-essence sector $\mS\supset\mathcal{K}(Q^2)$. If we choose $\mathcal{K}=c_1Q^2+c_2Q^4$ with $c_{1,2}$ some constants, its scalar field equivalent recovers the cubic Galileon ghost condensate that has been shown to provide a good fit to cosmological data in \cite{Peirone:2019aua}.

An important remark worth making at this point is that the TDiff symmetric teleparallel Horndeski terms that we have constructed in this section permit to go beyond the class of TDiff theories usually considered in the literature. For instance, the general TDiff theory analysed in \cite{Blas:2011ac} can be extended with the Horndeski TDiff terms obtained here without spoiling the second order nature of the field equations, thus opening a novel path for TDiff theories that remains unexplored. An obvious further extension are the TDiff symmetric teleparallel beyond Horndeski theories whose scalar field equivalents reproduce the theories constructed in e.g. \cite{Zumalacarregui:2013pma,Gleyzes:2014dya}.

Let us finally make a semantic observation. While the Lagrangians we considered are a subclass of the original Horndeski Lagrangians, the term symmetric teleparallel Horndeski has been used in Ref. \cite{Bahamonde:2022cmz} for more arbitrary Lagrangians which result in second order equations of motion though usually propagate more than an extra scalar degree of freedom besides the graviton.

\section{Matter couplings}
\label{Sec:MC}

Thus far, we have restricted to the pure gravitational sector, but a natural and pertinent question is how matter fields will couple to gravity in this framework. For a thorough discussion of matter couplings in teleparallel theories we will refer to \cite{BeltranJimenez:2020sih}. As was shown there, the symmetric teleparallel minimal coupling
($\partial_\mu \rightarrow \nabla_\mu$) is equivalent to
the metrical coupling ($\partial_\mu \rightarrow D_\mu$, which would be the minimal coupling in the conventional pseudo-Riemannian formulation of GR), so either way the resulting theory will be an equivalent GR with an extra
uncoupled scalar field plus the global degree of freedom $\lambda_0$. Effectively, matter fields only couple to the metric, but not to the independent connection.

More interesting may be to consider {\it non-minimal couplings} in the matter sector so that the matter fields cease being oblivious to the connection. In the symmetric teleparallel framework, it seems natural to introduce couplings to the connection through the non-metricity. However, this must be done with care not to introduce undesired new degrees of freedoms that could jeopardise the good properties of the gravitational sector, i.e., the absence of ghosts. Our guiding principle for this task can be the enhanced symmetry (i.e. TDiffs) of the gravitational sector that we should also respect in the matter couplings and this is trivially achieved if the non-minimal couplings are introduced through the non-metricity trace $Q_\alpha$, which is the building block for the construction of the gravitational sector as well. It is not difficult to see that these couplings will simply amount to introducing derivative interactions with the scalar field of the equivalent description. Thus, we could for instance couple the matter sector through a disformal metric defined by
\be
\tilde{g}_{\mu\nu}(Q_\alpha)=A(Q_\alpha Q^\alpha)g_{\mu\nu}+B(Q_\alpha Q^\alpha)Q_\mu Q_\nu
\label{eq:disformalQ}
\ee
with $A$ and $B$ some arbitrary functions. It should be clear that the equivalent scalar theory will feature a coupling of the scalar to the matter fields through the metric
\be
\tilde{g}_{\mu\nu}(\partial_\alpha \varphi)=A(X)g_{\mu\nu}+B(X)\partial_\mu \varphi\partial_\nu \varphi.
\ee
More explicitly, if we start from an action $\mS=\mS[g_{\mu\nu},\Psi]$ for the matter fields $\Psi$, we can introduce couplings by promoting $\mS[g_{\mu\nu},\Psi]\to\mS[\tilde{g}_{\mu\nu}(Q_\alpha),\Psi]$ and these couplings will eventually result in derivative interactions to the scalar field so the final action will be $\mS=\mS[\tilde{g}_{\mu\nu}(\partial_\alpha\varphi),\Psi]$. The resulting theory will then fit within Bekenstein's disformal couplings \cite{Bekenstein:1992pj}.

However, the disformal coupling is only a specific geometrical way of adding couplings to matter. In general, we can introduce arbitrary couplings to the non-metricity trace $Q_\alpha$, not necessarily via a disformal metric, that will give rise to derivative couplings to the scalar field of the equivalent formulation. In fact, we can even introduce matter couplings to derivatives of the non-metricity in an analogy with the pure gravitational couplings analysed in Sec. \ref{Sec:TDiffHorndeski}. Introducing couplings to matter in this way will trivially preserve all the symmetries (both local and global) already present in the gravitational sector. The generated interactions have a formal resemblance to the natural couplings that arise in higher dimensional braneworld scenarios where the Goldstone bosons associated to the breaking of the fifth-dimensional translations couple derivatively to the fields living on the brane. Thus, much of the phenomenology for those scenarios will also exist in the TDiff symmetric teleparallel theories (see e.g. \cite{Cembranos:2016jun} and references therein).\footnote{The discussed interactions through the shift-symmetric disformal metric can also be obtained from consistent couplings to the total energy-momentum tensor \cite{BeltranJimenez:2018tfy}.}

In order to illustrate the type of matter couplings that we have discussed, we shall consider a $U(1)$ gauge field $A_\mu$. For this field, preserving the gauge symmetry, we can introduce the following couplings to the non-metricity:
\begin{align}
\mS=\int\diff^4x\sqrt{-g}\Big[&\mathcal{A}(Q^2)F_{\mu\nu}F^{\mu\nu}+\mathcal{B}(Q^2)Q^\mu Q^\nu F_{\mu\alpha}F_{\nu}{}^{\alpha}\nonumber\\
&+\mathcal{G}_3(Q^2)\tilde{F}^{\mu\alpha}\tilde{F}^{\nu}{}_{\alpha}D_\mu Q_\nu
\nonumber\\
&+\mathcal{G}_4(Q^2)\tilde{F}^{\mu\alpha}\tilde{F}^{\nu\beta}D_\mu Q_\nu D_\alpha Q_\beta\Big]\,,
\label{eq:gaugecoupling}
\end{align}
where $F_{\mu\nu}=\partial_\mu A_\nu-\partial_\nu A_\mu$ is the field strength of the gauge field, $\tilde{F}^{\mu\nu}$ its Hodge dual and $\mathcal{A}$, $\mathcal{B}$, $\mathcal{G}_3$ and $\mathcal{G}_4$ arbitrary functions. The first line in \eqref{eq:gaugecoupling} corresponds to the ultra-local couplings to the non-metricity and can be recast as an interaction through a disformal metric like 
the one in \eqref{eq:disformalQ} with appropriate relations between $A,B$ and $\mathcal{A},\mathcal{B}$. The second and third lines in \eqref{eq:gaugecoupling} involve derivatives of the non-metricity trace $Q_\alpha$ and cannot, in general, be associated to a disformal metric.\footnote{Specific couplings of this type do admit however a geometrical origin from e.g. higher dimensional setups.} These are the matter sector analogous of the purely gravitational derivative interactions of Sec. \ref{Sec:TDiffHorndeski} as mentioned above. Following once again the procedure to integrate out the Stueckelberg fields and obtain the scalar field equivalent, the scalar will couple to the gauge field as follows:
\begin{align}
\mS=\int\diff^4x\sqrt{-g}\Big[&\mathcal{A}(X)F_{\mu\nu}F^{\mu\nu}+\mathcal{B}(X)\partial^\mu \varphi \partial^\nu\varphi F_{\mu\alpha}F_{\nu}{}^{\alpha}\nonumber\\
&+\mathcal{G}_3(X)\tilde{F}^{\mu\alpha}\tilde{F}^{\nu}{}_{\alpha}D_\mu D_\nu\varphi
\nonumber\\
&+\mathcal{G}_4(X)\tilde{F}^{\mu\alpha}\tilde{F}^{\nu\beta}D_\mu D_\nu\varphi D_\alpha D_\beta\varphi\Big]\,
\label{eq:gaugevarphicoupling}
\end{align}
with appropriate redefinitions of the coupling functions. Some comments are in order. Firstly, the reader might bring back the objection that we are employing the Levi-Civita covariant derivative $D_\mu$ so we are somehow abandoning the symmetric teleparallel framework. It is important however to bear in mind that this can be regarded as a notational convenience. Let us recall that the symmetric teleparallel connection can be expressed as 
\be
\Gamma^\alpha{}_{\mu\nu}=\{^\alpha_{\mu\nu}\}+L^\alpha{}_{\mu\nu}\,,
\ee
where $\{^\alpha_{\mu\nu}\}$ is the Levi-Civita connection of the metric and $L^\alpha{}_{\mu\nu}=\frac{1}{2}Q^\alpha{}_{\mu\nu} - Q_{(\mu\nu)}{}^\alpha$. Thus, we can always express a metric covariant derivative as a symmetric teleparallel derivative plus non-metricity terms $D=\nabla-L$ and this would allow as to write the matter couplings in the symmetric teleparallel framework. The second important comment is that the interactions to derivatives of the non-metricity trace generate couplings to second derivatives of the scalar field equivalent \eqref{eq:gaugevarphicoupling} with the associated hazard of generating Ostrogradski instabilities. We were careful with the precise form of the interactions to derivatives of the non-metricity trace
$Q_\alpha$ in \eqref{eq:gaugecoupling} to guarantee that no Ostrogradski ghosts are re-introduced from the matter couplings because the interactions to the dual of the field strength featured in \eqref{eq:gaugevarphicoupling} precisely correspond to healthy interactions between a gauge field and second derivatives of a scalar field (see e.g. \cite{Deffayet:2010zh,BeltranJimenez:2016rff}). Perhaps the simplest way to see that no higher than second order derivatives are generated is by noting that $\tilde{F}^{0i}$ gives the magnetic part of the electromagnetic field, which carries no time derivatives (the dynamical sector is carried by the electric piece).  It is then not difficult to convince oneself that the interactions in the second and third lines of \eqref{eq:gaugevarphicoupling} do not contain more than two time-derivatives, so that they will not produce higher than second order derivative terms in the field equations. This does not occur for instance for the interaction $F^{\mu\alpha}F^{\nu}{}_{\alpha}D_\mu D_\nu\varphi$ which does feature higher order derivatives in the equations of motion. Finally, let us notice that, since $Q_\alpha$ is a pure gradient in the symmetric teleparallel geometry, we have that $\nabla_{[\mu}Q_{\nu]}=D_{[\mu}Q_{\nu]}=0$ so we only need to consider the symmetric part of the covariant derivatives in the construction of the couplings to matter.

Slightly deviating from the symmetric teleparallel realm, one can be more general and couple the matter fields to the combination $g/\Lambda^2$ (without derivatives) that respects the desired symmetries and will also admit a scalar equivalent description. This is of course still Diff-invariant and preserves the additional TDiff symmetry.\footnote{These couplings have been considered in the context of TDiff theories in e.g. \cite{Maroto:2023toq,Jaramillo-Garrido:2023cor,Maroto:2024mkx}.} These couplings however break the global dilation/shift symmetry and including them would also require adding some general potential $V_{\text{TDiff}}(g/\Lambda^2)$ in the gravitational sector. The absence of the global symmetry in these interactions are likely to lead to a worse behaviour against quantum corrections and might require some fine-tuning. The reason for this is that the global symmetry will help to prevent quantum corrections that would renormalise the leading order operators, thus spoiling the predictability of the theory as an effective field theory. Since these interactions are arguably beyond the {\it natural} symmetric teleparallel framework, and they can be easily forbidden by appealing to the global dilation symmetry, we can safely leave them out.

\section{Discussion}
\label{Sec:Discussion}

The purpose of this note has been to show how the symmetric teleparallel framework allows for a class of ghost-free theories where the full Diffs that guarantee the absence of pathologies for the STEGR is broken down to its TDiff subgroup. For these theories, we have been able to fully integrate out the Stueckelberg fields of the symmetric teleparallel connection to show an equivalence with a family of scalar field theories with a potential. The remarkable feature of these theories is that the potential of the scalar field contains a parameter that corresponds to an integration constant that arises in a similar manner to the cosmological constant in unimodular gravity.  After discussing some properties of the equivalent theory, we have shown how to introduce couplings to matter as well as non-minimal couplings.  Apart from their theoretical consistency, the minimally coupled theories are identical to GR in the weak field limit having also the same PPN parameters and differing from other gravity theories featuring an additional scalar field such as Brans-Dicke, $f(R)$ or $f(G)$. Moreover, they  can recover the standard cosmological evolution in the $\phi=const$ and $\lambda_0=0$ case, so that they are also viable alternatives to Einstein gravity from an observational point of view. In other words, the theory contains solutions that are identical to those of GR so that we can be as close as desired to the standard evolution. On the other hand, it is interesting to notice that these theories will naturally come equipped with screening mechanisms (see e.g. \cite{Joyce:2014kja} for a review on scalar field theories featuring screening mechanisms) that exhibit the novelty of being governed by a global degree of freedom for some solutions.

It is worth to point out that the metric teleparallel admits an analogous (manifestly ghost-free) path to generate extensions of the MTEGR. The TDiff symmetry in that geometrical framework can be implemented by introducing terms depending on the determinant of the vierbein. One could then translate our findings to that framework and generate equivalent theories formulated in metric teleparallel geometries. However, while the symmetric teleparallel geometrical set-up admits a natural formulation in terms of the fundamental geometrical object, namely, the non-metricity, the analogous formulation in the metric teleparallel framework appears less natural to formulate in terms of the torsion (if possible at all). On the other hand, one could also attempt to construct theories along the lines suggested here for the general teleparallel geometry where the connection is only subject to be flat, but it is otherwise free, i.e., the reference frames do not need to be integrable, as in the symmetric teleparallel geometry, nor Lorentzian, as in the metric teleparallel geometry. The relevant symmetry group for the general equivalent is a full GL(4,$\mathbb{R}$) symmetry so there is more room for soft relaxations of the GR equivalent symmetries that could accommodate pathology-free theories. This program was already initiated to some extent at linear order in \cite{BeltranJimenez:2019odq} and those results can serve to select the potential ghost-free candidates.

We will conclude by emphasising that the symmetric teleparallel geometries provide a very natural place to formulate TDiff theories and our results open a new pathology-free avenue for exploring phenomenological applications as in e.g. cosmology or black hole physics.

\vspace{0.5cm}
{\bf Acknowledgments:}
J.B.J. was supported by the Project PID2021-122938NB-I00 funded by the Spanish “Ministerio de Ciencia e Innovaci\'on" and FEDER “A way of making Europe”. T.S.K. was supported by the CoE TK202 “Foundations of the Universe” funded by the Estonian Research Council. AGBM and ALM were supported by the MICIN (Spain) Project No. PID2022-138263NB-I00 funded by MICIU/AEI/10.13039/501100011033 and by ERDF/EU.

\bibliography{GhostFreeSymTel}

\end{document}